\documentclass[letterpaper,twocolumn,
prl 
,superscriptaddress
]{revtex4}

\usepackage{graphicx}
\usepackage{amssymb}
\usepackage{amsmath}
\usepackage{color}
\usepackage{comment}

\newcommand{\vect}[1]{{\mathbf #1}}

\newcommand{\bra}[1]{\left\langle{#1}\right|}
\newcommand{\ket}[1]{\left|{#1}\right\rangle}

\begin{document}


\title{Trimers, molecules and polarons in imbalanced atomic Fermi gases}

\author{Charles J. M. Mathy}
\affiliation{Department of Physics, Princeton University, Princeton,
New Jersey 08544, USA}

\author{Meera M. Parish}
\affiliation{Department of Physics, Princeton University, Princeton,
New Jersey 08544, USA} %
\affiliation{Cavendish Laboratory, JJ Thomson Avenue, Cambridge,
 CB3 0HE, UK}

\author{David A. Huse}
\affiliation{Department of Physics, Princeton University, Princeton,
New Jersey 08544, USA} %

\date{\today}

\begin{abstract}
We consider the ground state 
of a single ``spin-down'' impurity atom interacting attractively with a
``spin-up'' atomic Fermi gas.
By constructing variational wave functions for polarons, molecules
and trimers, we perform a detailed study of the 
transitions between each of these dressed bound states as a function of mass
ratio $r=m_\uparrow/m_\downarrow$ and interaction strength. 
We find that the presence of a Fermi sea \emph{enhances} the stability of the $p$-wave trimer, which can be
viewed as a Fulde-Ferrell-Larkin-Ovchinnikov (FFLO) molecule that has bound an additional majority atom.
%
For sufficiently large $r$, we find that the 
transitions lie outside the region of phase 
separation in 
imbalanced Fermi gases and should thus be
observable in experiment, unlike the well-studied equal-mass case. 
\end{abstract}


%

\maketitle

The spin-imbalanced Fermi gas has received much attention recently 
owing to its elegant realization in ultracold atomic gases. The
ability to tune both the interspecies interaction and the spin
polarization has allowed cold-atom experiments to access the rich
phase diagram of the spin-imbalanced system~\cite{atom_expts}.
%
However, as experiments become ever more precise, a challenge for
theory is to go beyond conventional mean-field approaches and
accurately determine the existence of exotic phases in the regime of
strong correlations.

One approach that has proven useful at high polarizations is to
consider the problem of a \emph{single} spin-down impurity atom
immersed in a Fermi gas of spin-up
atoms~\cite{chevy2006_2,lobo2006,prokofiev2008}. Such a scenario is
just one example of the canonical ``polaron'' problem, the solution
of which is used to construct the low-energy behaviour of the many-body system. 
%
Moreover, this limit of full polarization contains some of the
critical points of the full zero-temperature phase diagram, e.g., it
features the tricritical point that marks the existence of the
spatially-homogeneous superfluid phase for \emph{all}
polarizations~\cite{parish2007}.  Thus, an analysis of the
high-polarization limit allows one to characterize parts of the
topology of the whole phase diagram.

An important feature of the single-impurity problem is that it can
exhibit binding transitions where the impurity changes its
statistics and/or effective mass for a sufficiently strong
attractive interaction.  Thus far, the focus has been on equal
masses $m_\downarrow = m_\uparrow$, where it has been shown that the
impurity undergoes a first-order quantum phase transition from a
polaron (an impurity dressed with particle-hole excitations) to a
molecule (an impurity bound to a single 
majority fermion)~\cite{prokofiev2008}.
Here we demonstrate that a richer variety of phases
can be obtained when the masses are unequal, with $m_\downarrow<m_\uparrow$.
In this case, the presence of a spin-up Fermi sea can favor 
a dressed molecule with \emph{nonzero} ground-state momentum, corresponding
 to the FFLO superfluid phase in the limit of extreme imbalance. However, 
we find that the FFLO molecule generally prefers to bind another majority
 atom to form a $p$-wave trimer at zero momentum. 
Thus, as one increases the density of majority spins from the limit of zero density, one finds that the
stability of the trimer is initially \emph{enhanced} by the Fermi sea. 


In this Letter, we map out the ground-state phase diagram for the
dressed impurity as a function of mass ratio $r =
m_\uparrow/m_\downarrow$ and interaction strength.
We also examine the decay rates out of the various phases, after quenching across a phase transition.
Lastly, we determine whether or not our dressed-impurity phase diagram is
thermodynamically stable in the 
highly-imbalanced Fermi gas, by
estimating the onset of phase separation in this limit using the
results of previous quantum Monte Carlo (QMC) simulations.
In contrast to the equal-mass case, we find that the polaron-trimer binding
transitions sit \emph{outside} of the phase-separated region and should
thus be experimentally accessible.

In the following, we use the Hamiltonian for a two-component
(we call the two species $\uparrow$, $\downarrow$) atomic Fermi gas interacting via a wide
Feshbach resonance:
\begin{align}
 H = & \sum_{\vect{k}\sigma} \epsilon_{\vect{k}\sigma}
 c^\dag_{\vect{k}\sigma}c_{\vect{k}\sigma}
 + \frac{g}{V}\sum_{\vect{k},\vect{k'},\vect{q}}
 c^\dag_{\vect{k}\uparrow}c^\dag_{\vect{k'}\downarrow}
 c_{\vect{k'}+\vect{q}\downarrow}c_{\vect{k}-\vect{q}\uparrow}~,
\end{align}
where $\epsilon_{\vect{k}\sigma} = \frac{\vect{k}^2}{2m_\sigma}$ (we
set $\hbar = 1$), $V$ is the system volume and $g$ is the strength
of the attractive contact interaction. The $s$-wave scattering
length $a_s$ is then obtained via the prescription $\frac{m_rV}{4\pi
a_s} = \frac{V}{g} +
\sum_{\vect{k}}^{\Lambda}\frac{1}{\epsilon_{\vect{k} \uparrow} +
\epsilon_{\vect{k}\downarrow}}$, where the `reduced mass'
$\frac{2}{m_r} = \frac{1}{m_\downarrow} + \frac{1}{m_\uparrow}$, and
$\Lambda$ is a
UV cutoff that can be sent to infinity at the end of the
calculation.
Note that since the $\downarrow$ impurity is distinguishable from the 
$\uparrow$ Fermi sea, our results in the single impurity limit are relevant to
both Fermi gases and Bose-Fermi mixtures.

We construct variational wave functions for the single 
impurity atom immersed in a 
Fermi sea by considering
different numbers of particle and particle-hole pair excitations
upon the Fermi sea. Previous studies have shown this approach to be
reasonably accurate~\cite{combescot2008}. The novelties of our work
are to thoroughly explore the case of unequal masses, to allow the
dressed impurity to have non-zero momentum, and to consider the
possibility of trimers as well as polarons and molecules. We also considered 
tetramers 
at angular momentum $L=1$, but they did not 
bind anywhere for the parameters we considered.

We adopt the following nomenclature for the different trial wave
functions: a subscript $n$ refers to a state with at most $n$
operators acting on the non-interacting Fermi sea, and a bracketed
momentum refers to the total momentum of the state. For example, the
polaronic state $P_3(\vect{Q})$ will contain the impurity atom, and
at most one particle-hole pair on top of the Fermi sea, with a total
momentum $\vect{Q}$.
%
%
In our explicit calculations we concentrate on states with at most
one particle-hole pair, since these are known to provide a good
approximation for the impurity energy when $m_\uparrow =
m_\downarrow$~\cite{combescot2008,combescot2009,mora2009,punk2009}
and going to higher order greatly increases the numerical effort.
%
Thus, the 
wave function for the polaron is~\cite{chevy2006_2,combescot2008}:
\begin{align}
|P_3(\vect{Q}) \rangle & = \  \alpha^{(\vect{Q})}
c^\dag_{\vect{Q}\downarrow} | FS \rangle  
+ \sum_{\vect{k},\vect{q}} \beta^{(\vect{Q})}_{\vect{k}\vect{q}}
c^\dag_{\vect{Q}+\vect{q}-\vect{k}\downarrow}
c^\dag_{\vect{k}\uparrow} c_{\vect{q}\uparrow} |FS\rangle
\end{align}
where $| FS \rangle$ is a Fermi sea of majority atoms, filled up to
momentum $k_{F\uparrow}$. 
The presence of a Fermi sea implies that the spin-up hole momentum
$|\vect{q}| \equiv q <k_{F\uparrow}$ and the spin-up particle
momentum $\vect{k}$ satisfies $k_{F\uparrow} < k < \Lambda$.
%
Likewise, for the molecule, the wave function 
is~\cite{combescot2009,mora2009,punk2009}:
\begin{align}
\notag %
|M_4(\vect{Q})\rangle & = \sum_{\vect{k}}
\gamma^{(\vect{Q})}_{\vect{k}}c^\dag_{\vect{Q}-\vect{k}\downarrow}
c^\dag_{\vect{k}\uparrow} |FS\rangle \\
& + \sum_{\vect{k},\vect{k'},\vect{q}}
\delta^{(\vect{Q})}_{\vect{k}\vect{k'}\vect{q}}
c^\dag_{\vect{Q}+\vect{q}-\vect{k}-\vect{k'}\downarrow}
c^\dag_{\vect{k}\uparrow} c^\dag_{\vect{k'}\uparrow}
c_{\vect{q}\uparrow} |FS\rangle~.
\label{eq:M4}
\end{align}
Finally, for the trimer, we set $\vect{Q}=0$ and approximate the wave function as being insensitive to the hole momentum $\vect{q}$ by using its value at $\vect{q} = 0$ (this approach has been used for the polaron~\cite{combescot2008} and molecule~\cite{mora2009,punk2009}, and can be shown to give an upper bound to the binding energy):
\begin{align}\notag
&|T_5(0) \rangle = \sum_{\vect{k_1},\vect{k_2}}
\tau_{\vect{k_1}\vect{k_2}}
c^\dag_{-\vect{k_1}-\vect{k_2}\downarrow}
c^\dag_{\vect{k_1}\uparrow} c^\dag_{\vect{k_2}\uparrow} |FS \rangle \ + 
\\ 
& \sum_{\vect{k_1},\vect{k_2},\vect{k},\vect{q}} \eta_{\vect{k_1},\vect{k_2},\vect{k},0} c^\dag_{\vect{q}-\vect{k_1}-\vect{k_2}-\vect{k}\downarrow}
c^\dag_{\vect{k_1}\uparrow} c^\dag_{\vect{k_2}\uparrow} c^\dag_{\vect{k}\uparrow} c_{\vect{q}\uparrow} |FS\rangle~.
\end{align}
Here, we set the total angular momentum to be $L=1$ and choose
$L_z=0$, in which case the bare part is
\begin{align}\notag
\tau_{\vect{k_1}\vect{k_2}} &  ={ \hat
k_1\cdot\hat z}F(k_1, k_2,{\hat k_1\cdot\hat k_2})  - {\hat
k_2\cdot\hat z}F(k_2, k_1,{\hat k_1\cdot\hat k_2})
\end{align}
with any function $F$.
%
Unlike the polaron and 
molecule, which both have $L=0$ in the
ground state, we find that this $L=1$ odd-parity trimer is always
the lowest energy state
throughout the portion of the $\{1/k_{F\uparrow} a_s, r\}$ phase
diagram where the trimer is stable.  We have not yet explicitly looked at trimer 
states with nonzero momentum, but, as far as we are aware, there is no indication that the trimer has a lower energy at $\vect{Q}\neq 0$.

In the limit $k_{F\uparrow} \rightarrow 0$, this $L=1$ trimer
becomes the 3-particle bound state in a vacuum, for which analytical
solutions have been found~\cite{trimer}. Here, it has been shown
that the trimer is bound relative to a molecule and an extra
particle for $r>r_{C1}\cong 8.17$. However, for $r>r_{C2}\cong
13.6$, the energy of the trimer is no longer finite in the limit
$\Lambda \rightarrow \infty$. This shows up as a
wave function with weight at increasingly high momenta as one
approaches $r_{C2}$ from below.
This critical $r_{C2}$ is independent of $k_{F\uparrow}$, since it
relies on high momenta. Thus, the results for the trimer are always
cutoff dependent once $r>r_{C2}$.
%

We solve for the ground-state energy of a given normalized wave function
$\ket{\psi}$ by taking $E = \bra{\psi} H \ket{\psi}$ and then
minimizing the energy $E$ with respect to the amplitudes 
$\alpha$, $\beta$, $\gamma$, $\delta$, $\tau$ and/or $\eta$. One can
generically sum over one of the momenta, and solve a self-consistent
equation for the resulting function, where symmetries play a crucial
role in simplifying the calculation. For example, for the trimer
with $L=1$ and $L_z=0$, we define $\frac{g}{V}\sum_{\vect{k_2}}
\tau_{\vect{k},\vect{k}_2} = G(k) \hat k \cdot \hat z$, and then
derive a self-consistent equation for $G$.

\begin{figure}
\centering
\includegraphics[width=0.9\linewidth]{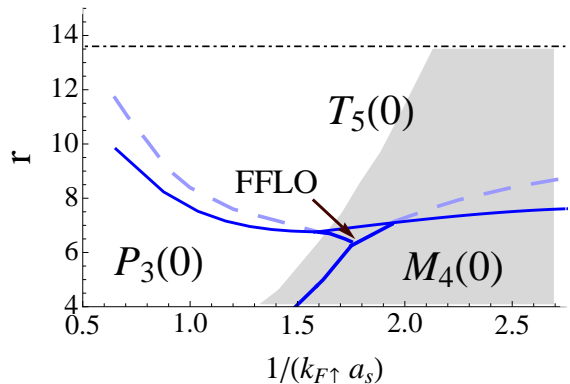}
\caption{(Color online) The ground-state phase diagram as a function
of mass ratio $r$ and interaction strength $1/k_{F\uparrow} a_s$ for
the polaron ($P_3$), molecule ($M_4$) and trimer ($T_5$) wave
functions.
The FFLO region corresponds to $M_4$ with non-zero momentum;
the momentum of the FFLO molecule is approximately $k_{F\uparrow}$ at the $P_3(0)$-FFLO transition line and goes continuously to zero at the
FFLO-$M_4(0)$ transition line, coming in as a square root near the transition. 
The $T_5$-$M_4$ boundary (full [blue] line) approaches the 3-body transition $r_{C1}
\simeq 8.17$ in the limit $1/k_{F\uparrow} a_s \rightarrow \infty$,
as expected.
Above the dashed-dotted ($r=r_{C2}$) line, the results for 
$T_5$ become cutoff-dependent, and are therefore no longer
universal.  
%
The shaded region marks where the system is unstable to phase
separation. 
The dashed (blue) line marks the transition line where a metastable polaron rapidly decays into a FFLO molecule which then decays into a trimer. 
}
\label{fig:P3T5M4}
\end{figure}

By determining the minimum energies of the wave functions $P_3(\vect
Q)$, $M_4(\vect Q)$, $T_5(0)$, we constructed an approximation to
the ground-state phase diagram of the impurity atom, as depicted in
Fig.~\ref{fig:P3T5M4}.
Here, we find that the polaron phase always has its lowest energy at
zero momentum. 
By contrast, a very small region in
Fig.~\ref{fig:P3T5M4} exists where the molecule has \emph{nonzero} momentum
in the ground state (a FFLO molecule).
%

The FFLO phase has a very natural interpretation in this limit:
having a molecule with $\vect Q=0$ requires the impurity atom to
have momentum $k>k_{F\uparrow}$, which is kinetically disfavored as
$r$ increases. Thus, a FFLO molecule appears once the impurity's
momentum drops below $k_{F\uparrow}$.
However, as $r$ is increased farther, it quickly becomes favorable
for the FFLO molecule to bind yet another 
majority fermion and form a
$p$-wave trimer at zero total momentum.  Indeed, at large enough
$1/(k_{F\uparrow}a_s)$, the transition is directly from the $\vect
Q=0$ molecule to the trimer, without any intervening FFLO phase. 
A surprising result is that as one approaches unitarity, the Fermi surface 
actually favors the trimer phase over the molecule phase and we have a 
direct transition from polaron to trimer. 
We expect this to be a robust result since our approximation for $T_5$ underestimates the trimer binding energy. 
%

How are these phases distinct from one another?
In the polaron phase, the ground state wavefunction (including
dressing with an arbitrarily large number of particle-hole pairs)
has a nonzero weight $Z=|\alpha^{(\vect Q)}|^2$ for the ``bare''
impurity atom. This weight appears to remain nonzero in the
thermodynamic limit only at momentum $\vect Q=0$: the effective mass
$m^*$ of the polaron appears to be positive whenever the polaron is
the ground state in our phase diagram, thus a polaron with a nonzero
$Q$ has higher energy by $Q^2/(2m^*)$ and is unstable to emitting a
particle-hole pair and scattering to lower momentum.
%
We may consider this weight $Z$ to be the ``order parameter'' for
the polaron phase.

In the molecule phase, we have $Z=0$, but a fully-dressed ground state will
have a nonzero weight $Z_M=\sum_{\vect{k}}
|\gamma^{(\vect{Q})}_{\vect{k}}|^2$ for the bare molecule at one or
more values of $\vect Q$. The FFLO molecule phase is where this
nonzero weight appears at $Q>0$. Thus again we can consider this
nonzero $Z_M$, as well as the momentum $Q$ where it occurs, as the
``order parameters'' of the molecule phases. Similarly, the trimer
phase is where the ground state has a nonzero weight 
$Z_T = \sum_{\vect{k_1}\vect{k_2}} |\tau_{\vect{k_1}\vect{k_2}}|^2$ 
for the bare trimer.

\begin{figure}
\centering
\includegraphics[width=0.8\linewidth]{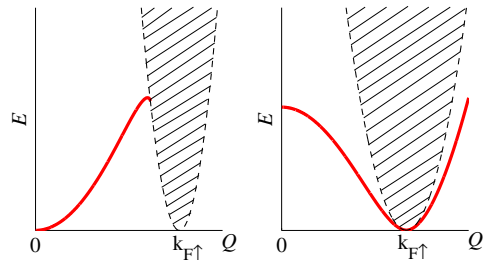}
\caption{(Color online) Schematics of two different scenarios for a
molecule unbinding into a polaron + particle. The solid (red) lines
represent the molecule dispersion $E(Q)$ and the shaded regions
correspond to the polaron + particle (two-body) continuum.
When both the molecule and polaron have their minimum energies at
$\vect{Q}=0$ (left), the transition is first-order. However, we have
a continuous transition (where the bound state fully ``mixes''
with the continuum) when the molecule has ground-state momentum
$Q = k_{F\uparrow}$ (right).} \label{fig:Unbinding}
\end{figure}

Next, we examine the nature of the binding transitions for the
single impurity.
For the case of equal masses, it is known that the matrix
elements connecting the polaron and molecule 
go to zero at the transition point due to the transition involving a
``decay'' into at
least four particles~\cite{prokofiev2008}.  Thus, one can 
have a long-lived molecule (polaron) existing on the polaron
(molecule) side of the transition, with the lifetime $\Gamma^{-1} \sim (\Delta E)^{-9/2}$ diverging
faster than the inverse of the energy difference $\Delta E$ as one approaches
the unbinding transition~\cite{Bruun}. This first-order
binding transition in the presence of the Fermi sea occurs when the
momenta of the two phases do not differ by $k_{F\uparrow}$.
For example, when the molecule and polaron both have their minimum
energy at $\vect{Q} = 0$, the molecule at $\vect{Q} = 0$ cannot
simply unbind into a polaron at $\vect{Q} = 0$ plus a low energy
spin up particle at the Fermi momentum $k_{F\uparrow}$, without
violating momentum conversation. Instead, it must create a
particle-hole pair, thus leading to the 4-particle process mentioned
earlier and a first-order transition. Referring to
Fig.~\ref{fig:Unbinding}, this follows from the displacement of the
molecule dispersion from the polaron + particle continuum by
$k_{F\uparrow}$. The same considerations apply to the zero-momentum
trimer unbinding into a zero-momentum molecule plus a particle.

However, if the molecule has its energy minimum at $Q =
k_{F\uparrow}$, then it can simply unbind in a continuous fashion
into either a zero-momentum polaron plus a particle (see
Fig.~\ref{fig:Unbinding}), or a zero-momentum trimer plus a hole.
This behaviour also shows up in the decay rates of the excited states  
on either side of the transition, which we can estimate using a simple Fermi's golden rule like in Ref.~\cite{Bruun}. We find that the unbinding of a FFLO molecule into a trimer or polaron has decay rate $\Gamma \sim \Delta E$ as $\Delta E \to 0$, where, once again, $\Delta E$ is the energy difference between the two phases near the transition. Therefore, this is a marginal case where the lifetime $\Gamma^{-1}$ diverges as fast as the inverse of $\Delta E$. The decay of a trimer or polaron into a FFLO molecule is even faster, with $\Gamma$ being finite as $\Delta E \to 0$, since the momentum of the final-state FFLO molecule 
can lie anywhere on the Fermi surface 
and thus the phase space is enlarged. 
Finally, the transition from the zero-momentum molecule to the FFLO
molecule is not an unbinding transition and we find that $\vect Q$
generally moves continuously away from zero at this transition.

A direct transition from the trimer to the polaron and vice versa would be
first-order, since this is a three-body decay: the trimer (polaron)
``shedding'' two particles (holes). In this case, the decay rate is always 
$\Gamma \sim (\Delta E)^2$ close to the transition, similar to the behavior of a Fermi liquid quasiparticle. 
This implies that a quench at fixed $r$ from the polaron to trimer phase can lead to a metastable polaron that can become unstable to forming a FFLO molecule, as shown in Fig.~\ref{fig:P3T5M4}. If the difference between the FFLO and trimer energies is small here, then there may exist a metastable FFLO phase, a scenario which is worthy of further investigation.

In experiments on highly-imbalanced 
Fermi gases, one has a nonzero number
density $n_{\downarrow}$ of minority atoms, and the system must be stable against
phase separation in order for the single-impurity transitions to be
observable.
For equal masses, QMC predicts that the single-impurity transition
occurs in the phase-separated regime~\cite{pilati2008}, and,
indeed, the experimentally measured disappearance of the
polaron~\cite{schirotzek2009} agrees well with the onset of phase
separation between superfluid (SF) and normal (N) phases, rather
than with the polaron-molecule transition for a single impurity.
To estimate the onset of phase separation for general mass ratios,
we impose the following coexistence conditions on the pressures $P$
and chemical potentials $\mu$ in each phase: $P_{SF} = P_{N}$, $\mu^{SF}_\sigma = \mu^{N}_\sigma$, and $\mu_\downarrow^{N} = E_b$, 
%
%
where $E_b$ is the binding energy of the impurity (a polaron,
molecule or trimer) immersed in the 
Fermi gas.
%
%
%
Fortunately, for mass ratios $r \geq 1$, the superfluid is likely to
be unpolarized 
($n_{\uparrow}=n_{\downarrow}$) or only
weakly polarized
at this onset point, and thus we can exploit the QMC
equation of state for the unpolarized 
superfluid. At unitarity, this
equation of state
has been shown to be relatively insensitive to mass
ratio~\cite{unpolSF}, while in the BCS and BEC limits, the equation
of state is that of a weakly-attractive Fermi gas and a
weakly-repulsive Bose gas, respectively. Thus, we have pressure and
average chemical potential:
\begin{align}
P_{SF} & = \frac{2^{5/2}m_r^{3/2}}{15\pi^2}
\varepsilon_F^{5/2} \ g(1/k_Fa_s, r) \\
\mu_{SF} & \equiv \frac{\mu_\uparrow^{SF} + \mu_\downarrow^{SF}}{2}
= \varepsilon_F \ f(1/k_Fa_s,r)
\end{align}
where the functions $g(1/k_Fa_s, r)$ and $f(1/k_Fa_s,r)$ are
determined via numerical interpolation between the known limits,
while $k_F = (6 \pi^2 n^{SF}_\uparrow)^{1/3}$ and $\varepsilon_F =
k_F^2/2m_r$.

Clearly, the pressure and spin-up chemical potential in the
fully-polarized normal phase are known exactly:
\begin{align}
P_N & = \frac{2^{5/2}m_r^{3/2}}{15\pi^2} \frac{1}{1+r} \
\varepsilon_{F\uparrow}^{5/2}\\
\mu^N_\uparrow & = \frac{2}{1+r} \ \varepsilon_{F\uparrow}
\end{align}
where $\varepsilon_{F\uparrow} = (k_{F\uparrow})^2/2m_r$.
To determine $\mu^N_\downarrow$, we estimate the binding energy
$E_b$ using the $P_3$ wave function.
Note that at large $r$, where the polaron is no longer the ground
state, this $E_b$ will be an underestimate, and so the calculated
onset position $1/k_{F\uparrow}a_s$ will be lower than the exact
result in this limit.
By applying coexistence conditions, 
we arrive at the set of equations
\begin{align}
\left( \frac{k_{F\uparrow}}{k_F} \right)^5 & = (1+r) \ g(1/k_Fa_s,r) \\
g(1/k_Fa_s,r) & = (1+r)^{3/2} \left( \frac{f(1/k_Fa_s,r)}{1 +
(1+r)\frac{E_b}{\varepsilon_{F\uparrow}}} \right)^{5/2}
\end{align}
For equal masses, our 
calculation gives 
$1/k_{F\uparrow}a_s \simeq 0.75$, which agrees well with
fixed-node QMC calculations~\cite{pilati2008}.

This onset line determines the region of phase separation near
unitarity and this is plotted in Fig.~\ref{fig:P3T5M4}. For
sufficiently large $r$, we see that the polaron-trimer transition
extends well outside of the regime of phase separation, and thus
should be observable in the polarized gas. This allows one to investigate 
first-order binding transitions and the possible metastable states discussed earlier. 



A remaining question is what happens to the trimer phase when there is a
finite density of spin-down atoms. 
In one dimension, it is known that a trimer phase exists for $n_\downarrow/n_\uparrow
= 1/2$, provided the interactions are sufficiently
large~\cite{Orso2010}.
Here, we expect a Fermi liquid
of trimers for low densities $n_\downarrow/n_\uparrow \ll 1/2$. 
However, it is possible that the FFLO phase will eventually win over the trimer phase 
as $n_\downarrow/n_\uparrow$ is increased. 
One may also have a mixture of trimers, polarons and/or
molecules as we approach the single-impurity binding transition. 
%

%
Experimentally, one can explore our phase diagram using two different
atomic species with unequal masses or by
artificially increasing the effective mass of the spin-up atoms with
a spin-dependent optical lattice. $^{40}$K-$^{6}$Li mixtures are the
favored choice, and, in Fig.~\ref{fig:P3T5M4}, their mass ratio
($r\simeq 6.7$) just touches the bottom parts of the trimer and FFLO regions.
%
%
Since each phase is characterized by a different effective mass, one
way of distinguishing these phases is to determine the effective
mass using the low-lying compression modes of the gas, as in
Ref.~\cite{Nascimbene}.
Another possibility is to directly measure the ``order parameter'':
Ref.~\cite{schirotzek2009} has successfully measured $Z$ in the
polaron phase using RF spectroscopy, but it remains an open question
whether or not this method can be extended to study $Z_M$ or $Z_T$.
%


\acknowledgments
 The work at Princeton was supported under ARO Award
W911NF-07-1-0464 with funds from the DARPA OLE Program. MMP
acknowledges support from the EPSRC.


\end{document}